# Highly Accurate Local Pseudopotentials of Li, Na, and Mg for Orbital Free Density Functional Theory

Fleur Legrain, Sergei Manzhos[1]

Department of Mechanical Engineering, National University of Singapore, Block EA #07-08, 9 Engineering Drive 1, Singapore 117576

**Abstract**

We present a method to make highly accurate pseudopotentials for use with orbital-free density functional theory (OF-DFT) with given exchange-correlation and kinetic energy functionals, which avoids the compounding of errors of Kohn-Sham DFT and OF-DFT. The pseudopotentials are fitted to reference (experimental or highly accurate quantum chemistry) values of interaction energies, geometries, and mechanical properties, using a genetic algorithm. This can enable routine large-scale ab initio simulations of many practically relevant materials. Pseudopotentials for Li, Na, and Mg resulting in accurate geometries and energies of different phases as well as of vacancy formation and bulk moduli are presented as examples.

## 1. Introduction

Realistic (i.e. including the effects of interfaces, microstructure etc.) and accurate simulations of materials at the atomistic scale require ab initio calculations on system with sizes of $10^4$-$10^7$ atoms. For decades, ab initio materials simulations have been dominated by Kohn-Sham (KS-) DFT (Density Functional Theory) [1] as the only practical option. DFT expresses the ground state energy of any system with $N_e$ electrons as a functional of its electron density $\rho(\boldsymbol{r})$: $E =$

[1] Corresponding author. Tel: +65 6516 4605. Fax: +65 6779 1459.
E-mail: mpemanzh@nus.edu.sg

$E[\rho(r)]$. In KS-DFT, $\rho$ is computed as $\rho = \sum_{i=1}^{N_e} \rho_i(r)$, where $\rho_i = |\phi_i(r)|^2$. The $\phi_i$ (called orbitals) can be found from the KS equation

$$-\frac{1}{2}\nabla^2 \phi_i + V_{eff}(r|\rho)\phi_i = \epsilon_i \phi_i. \quad (1)$$

Then

$$E_{KS} = -\frac{1}{2}\sum_{i=1}^{N_e}\int \phi_i^* \nabla^2 \phi_i dr + \int V(r)\rho(r)dr + \frac{1}{2}\iint \frac{\rho(r)\rho(r')}{|r-r'|}dr + E_{ion} + E_{XC}[\rho(r)]. \quad (2)$$

Here $V_{eff} = V(r) + V_H(r) + V_{XC}(r)$, where $V_H(r) = \int \frac{\rho(r')}{|r-r'|}dr'$, $V(r)$ is the Coulombic potential between the electron density and the nuclei, and $E_{ion}$ is the Coulombic potential between the nuclei. $V_{XC}$ is the (unknown) exchange-correlation (XC) potential, which ensures that the density computed from Eq. 1 is equal to the true electron density and is a functional derivative of the exchange-correlation energy $E_{XC}$, $V_{XC}(r) = \delta E_{XC}/\delta\rho(r)$. Multiple approximations to the XC term exist which provide acceptable accuracy of simulation of specific classes of materials or of specific properties. Different approximations are used in different applications. In KS-DFT, it is necessary to compute at least as many orbitals as there are electrons (or pairs thereof). Further, the KS equation (Eq. 1) must be solved iteratively (as $V_{eff}$ depends on $\rho$). The solution involves large basis set expansions of the orbitals. As a result of these, the scaling of CPU cost of KS-DFT is $O(N_e^3)$. Simulations are therefore limited to very small model systems, typically of dozens to thousands of atoms and cannot properly model mechanical properties (microstructure-driven) or large organic systems or interfaces. While heroic KS-DFT calculations are sometimes done on up to $10^6$ atoms [2], they are of limited use for materials design, as they cannot be routinely repeated for screening purposes, and they are difficult to verify. There is need for *routine* ab initio calculations on large systems.

Orbital-free DFT (OF-DFT) [3] is a promising alternative that enables large-scale computations at reasonable computational costs. It computes $E[\rho(r)]$ without computing $\phi_i$, resulting in orders of magnitude faster calculations and significant memory savings [4-6]. The scaling of OFDFT is near linear with system size. With OFDFT, systems with $O(10^5)$ atoms can be modeled using modest computational resources (a desktop computer), and more than $O(10^6)$ of atoms can be modeled using large-scale parallel computing environment. The last three terms of $E_{KS}$ (Eq. 2) can be used in OF-DFT, with the same approximations for $E_{XC}$ as in KS-DFT. However, the first term cannot be adopted from KS-DFT, and has to be approximated with different kinetic energy functionals (KEF), $T_s[\rho(r)]$. Several KEF approximations exist. Usually, the KEF is written as [7,8]

$$T_s[\rho(r)] = T_{TF}[\rho] + T_{vW}[\rho] + T_{NL}[\rho] \qquad (3)$$

where $T_{TF}[\rho] = \frac{3}{10}(3\pi)^{2/3} \int \rho^{5/3}(r)dr$, $T_{vW}[\rho] = \frac{1}{8}\int \frac{|\nabla \rho|^2}{\rho}dr$, and $T_{NL}[\rho] = C \iint \rho^a(r)\omega(\zeta(r,r'),|r-r'|)\rho^b(r')drdr'$ with $C, a$ and $b$ being model-dependent. Encouraging results have been obtained for metals with the WT functional [9, 10] where $\zeta(r,r') = const$ and with the WGC functional [11], where $\zeta(r,r') = \left(\frac{1}{2}\left([3\pi^2\rho(r)]^{\frac{\gamma}{3}} + [3\pi^2\rho(r')]^{\frac{\gamma}{3}}\right)\right)^{1/\gamma}$. For semiconductors, the HC (Huang-Carter) functional [12, 13] has been shown to perform better; there, $\zeta(r,r') = [3\pi\rho(r)]^{1/3}\left(1 + \lambda\frac{|\nabla\rho|^2}{\rho^{4/3}}\right)$. Several other functionals are available [14-16]. It is only recently that researchers have begun mapping the performance of different KEFs for different classes of materials. We note that KS-DFT started being widely used by the materials community when different XC functionals emerged, which were benchmarked and shown to provide acceptable accuracy for specific classes of materials or for specific properties (e.g. PBE [17] for solids or B3LYP [18] for molecules). A similar process is underway for OF-DFT [19]. This process is inhibited by the fact that *pseudopotentials (PP) which can be used with OF-DFT are not available for most elements of the periodic table*.

DFT simulations of system with $10^2$-$10^3$ atoms, and especially, plane-wave based calculations on periodic system (which is the approach most often adopted in materials modeling), are enabled by the use of PPs, which replace the ion and core electrons with an effective potential. The ionic potential which enters Eq. 2 and which is simply $-\frac{Z_A}{r}$ (where $Z_A$ is atomic number) is replaced with a function $v(r)$ which is continuous and which asymptotically behaves as $-\frac{Z}{r}$ for large $r$, where $Z$ is a valence charge. This allows considering explicitly only valence (and sometimes selected core) electrons and results in smoother electron density, which can be expanded in plane waves at reasonable cost. Highly accurate PPs used in KS-DFT are non-local in that electrons of different angular momenta (e.g. *s, p, d, f*) feel different potentials [20-22]. OF-DFT relies on local PPs, which are in general less accurate [13, 23-29]. Local PPs for OF-DFT are usually built by matching KS-DFT calculations with local and non-local PPs, either for isolated atoms or in bulk [13, 23]. This procedure has important disadvantages: (i) the obtained PPs are benchmarked to KS-DFT with non-local PPs, and as a result, they fully incorporate their errors. This guarantees that the accuracy of OF-DFT simulations with such PPs will always be *worse* than KS-DFT. (ii) PPs for KS-DFT need to be constructed for specific XC functionals (often application-dependent) because DFT errors with existing functionals are still substantial. A PP therefore needs to be tuned to a specific DFT setup. When tuning a local PP to KS-DFT with a non-local PP, one does not match the PP to the computational method with which the local PP is to be used (i.e. OF-DFT with a specific KEF). This means that the quality of OFDFT simulations may be *much worse* than that of KS-DFT.

To mitigate this compounding of errors, we propose a new procedure to build local PPs, in which the PP is tuned by using OF-DFT rather than KS-DFT, to reproduce a series properties, which include structural parameters and energies. We posit that such PPs will perform well for materials of similar kind, and if tuning is thorough, they will provide quantitative accuracy for a class of materials, which has been achieved with OF-DFT only sporadically [13]. Specifically, we parametrize the PP function and use a genetic algorithm to fit the parameters, in an automated feedback loop, to the reference geometries, energies, and mechanical properties. While parametrization of local PPs has been used before [24-28], it was not used with OF-DFT, and such a feedback loop has never been implemented to produce PPs matched to a given OF-DFT approximation (e.g. XC+KEF). We apply this approach to build highly accurate local pseudopotentials for Li, Na, and Mg, which reproduce lattice parameters and relative energies of different crystal phases, as well as the vacancy formation energies and the bulk moduli. We compare our results for Li and Mg to the existing highly accurate local PP for Mg by Carter et al. [13]. We also show that our method can be used to fit simultaneously atomic and bulk properties, which is important for applications where cohesive and insertion energies are important, such as electrochemical batteries [19, 30]. The compromising nature of the resulting PP highlights the need for KEFs which are accurate for both isolated atoms and bulk systems.

## 2. Methods

The pseudopotentials in real space were parametrized as

$$v(r) = -\frac{Z}{r}\tanh^{\frac{1}{\alpha}}(ar^\alpha) + B\exp\left(-\left(\frac{r}{b}\right)^\beta\right) + C\exp\left(-\left(\frac{r-r_c}{c}\right)^\gamma\right) + D\exp\left(-\left(\frac{r-r_d}{d}\right)^\delta\right) \quad (4)$$

where the values of $a$ and $\alpha$ were preset (not changed during the fit). We used atomic units, in which case $a = \alpha = 1$ was used. $Z$ is the valence charge of the atom (1 for Li and Na and 2 for Mg). The functional form of Eq. 4 ensures the correct asymptotic behavior $(-\frac{Z}{r})$ at large $r$. The 11 parameters $B, b, \beta, C, r_c, c, \gamma, D, r_d, d,$ and $\delta$ were fitted to reference values of lattice vectors $\{a_1, a_2, a_3\}$, differences in cohesive energies $E$ of different phases, and the bulk moduli $B_0$ of the most stable phases (*bcc* for Li and Na and *hcp* for Mg). The vacancy formation energy $E_{vac}$ was also computed although not included in the fit, for comparison with measured values. The fit minimized the objective function

$$\epsilon = w_{LA}\epsilon_{LA} + w_E\epsilon_E + w_{B_0}\epsilon_{B_0} \qquad (5)$$

where

$$\begin{aligned}
\epsilon_{LA} &= \epsilon_{LA}^{hpc} + \epsilon_{LA}^{fcc} + \epsilon_{LA}^{bcc} \\
\epsilon_{LA}^X &= |a_1^c - a_1^{ref}| + |a_2^c - a_2^{ref}| + |a_3^c - a_3^{ref}| \\
\epsilon_E &= |(E_1^c - E_{RT}^c) - (E_1^{ref} - E_{RT}^{ref})| + |(E_2^c - E_{RT}^c) - (E_2^{ref} - E_{RT}^{ref})| \\
\epsilon_B &= |B_0^c - B_0^{ref}|
\end{aligned} \qquad (6)$$

The superscript $c$ is for values computed with OF-DFT and *ref* for reference values. Subscript *RT* denotes the most stable phase at normal conditions and subscripts 1 and 2 the other two phases (i.e. *fcc* and *hcp* for Li and Na, and *bcc* and *fcc* for Mg). The reference values for $B_0$ were 0K estimates based on experimental data. Bulk moduli were computed from the stresses computed at 1.05x and 0.95x of the equilibrium volume as

$$B_0^c = \frac{B_0^{0.95V} + B_0^{1.05V}}{2} \qquad (7)$$

where

$$B_0^{V \pm \Delta V} = \frac{|\sigma_{xx}| + |\sigma_{yy}| + |\sigma_{zz}|}{3 \times \frac{\Delta V}{V}} \qquad (8)$$

The objective function was minimized with a genetic algorithm programmed in Octave [31]. The population size was set to 10,000 and the number of generations to 1000. The default values for other parameters of Octave's GA function were sufficient to achieve close fits to reference values. We confirmed that gradient-based optimization following GA minimization did not result in any noticeable improvement. The weights $w_{LA}, w_E, w_{B_0}$ were chosen to make sure that all error components $\epsilon_{LA}, \epsilon_E, \epsilon_{B_0}$ are minimized.

OF-DFT calculations were performed with the plane wave code PROFESS 2.0 [32]. The PBE exchange-correlation functional was used [17]. The cutoff for the plane wave expansion of the density was 800 eV. The tolerance for the energy in the self-consistency cycle was $1 \times 10^{-6}$ eV, and forces were relaxed to below $1 \times 10^{-2}$ eV/Å. These settings provided converged values. We performed calculations with the WT [9, 10] and WGC [11] KEFs. The GA optimization was performed with WGC, and the properties also computed with the WT functional using the optimized PP, except for the Mg PP fit involving $E_{coh}$ which used the WT functional (see Section 3.3).

The fit is able to achieve smaller $\epsilon_{LA}$ and $\epsilon_E$ values than error bars typically associated with experimental values. For example, 0K lattice vectors and differences in cohesive energies without the effect of vibrations (which are computed by DFT here) are often estimated from finite temperature measurements on non-ideal (and vibrating) crystals. We therefore chose to fit to highly accurate electronic structure reference calculations on pure crystals at 0K which are in agreement with values deduces from experiments (see Tables 1-3), except for the bulk modulus $B_0$ for which experimental estimates are taken. Reference all-electron DFT calculations were performed with the FHI-AIMS code [33]. The same PBE exchange-correlation functional was used as in OF-DFT calculations [17]. The basis sets and integration grids were set to "really_tight" settings to approach the converged basis limit. The convergence criteria for the self-consistency cycle were $1 \times 10^{-7}$ eV for the energy and $1 \times 10^{-6}$ for the density. For bulk calculations, 15 *k*-points per dimension were used for all crystals. Optimizations were performed with a force tolerance of $1 \times 10^{-2}$ eV/Å. The computed cohesive energies for Li, Na, Mg are 1.67, 1.10, and 1.51 eV, respectively, in good agreement with available measurements [34]. Other computed parameters are given in Tables 1-3 and also agree with experimental data [35-38].

## 3. Results

### 3.1. Lithium

The results of the fit of a local PP of Li and the final PP parameters are given in Table 1. For comparison, properties computed with the available PP of Carter [39] are also given, as well the reference values. Both ours and the PP of Carter (for comparison) are plotted in Fig. 1. The fitted PP reproduces more accurately the lattice parameters as well as the relative energies of *bcc*, *fcc*, and *hcp* phases. The errors is energy differences of less than 0.001 eV are certainly much smaller than the accuracy of KS-DFT. The bulk modulus $B_0$ and the vacancy formation energy $E_{vac}$ (*not* fitted) are also reproduced more accurately. We obtain $B_0$=15.2 GPa vs. the reference value of 13.9 GPa and Carter PP value of 16.2 GPa. We obtain $E_{vac}$=0.45 eV vs. the reference value of 0.48 eV and Carter PP value of 0.69 eV. We note that by the choice of the weights $w_{LA}, w_E, w_{B_0}$ it is possible to improve $B_0$, albeit at the expense of $E_{vac}$, for example, a fit with an accurate $B_0$=14.1 GPa results in $E_{vac}$=0.3 eV. Pseudopotentials can therefore be produced which are more accurate for specific applications.

*3.2. Sodium*

The results for Na are given in Table 2 and the fitted PP is plotted in Fig. 1. The potential parameters are also given in Table 2. Geometric parameters, relative energies of *bcc*, *fcc*, and *hcp* phases and the vacancy formation energy are reproduced with very high accuracy. The accuracy of the bulk modulus is typical of that achieved with DFT [40, 41].

*3.3. Magnesium*

The results for Na are given in Table 3 and the fitted PP is plotted in Fig. 1. The potential parameters are also given in Table 3. Here also, geometric parameters, relative energies of *bcc*, *fcc*, and *hcp* phases and the vacancy formation energy are reproduced with very high and similar accuracy to that achieved with the PP of Carter [13]. Our bulk modulus of 35.9 GPa is somewhat closer to the reference value.

For Mg, we also attempted to produce a PP reproducing simultaneously bulk properties as well as the cohesive energy, which requires a single atom calculation. Existing KEFs, including the KEFs used here, are known to be in significant error for atomic and molecular properties [3, 42]. For example, the cohesive energy of Mg computed with Carter PP is 0.05 eV with the WT KEF, and the calculation with the WGC KEF did not converge. Our PP fitted to bulk properties does not fare any better (Table 3). Nevertheless, it might be advantageous to *effectively* include atomic properties even when relying on very approximate KEFs, by adjusting the pseudopotential. This would permit simulations where cohesive energies and defect formation energies with respect to vacuum state are important, such as doped materials or battery materials, specifically materials for Li, Na, and Mg ion batteries [19, 30, 43].

We are able to fit a PP which does reproduce $E_{coh}$ while maintaining an acceptable accuracy for all geometries and energies (see Table 3). In this fit, the values of $B_0$ and relative energies of *bcc* and *fcc* phases were *not* included and are computed a posteriori. The bulk modulus of 46.2 GPa is noticeably different from the reference value, while energy differences between phases, which are reproduced with within ~0.05 eV, remain within typical DFT accuracy [44]. It is therefore possible to effectively account for single-atom properties within the available OF-DFT setups. However, the differences in computed values of properties and in the PP shape (also shown in Fig. 1 as grey line, with PP parameters listed at the bottom of Table 3) between bulk-only fits and fits including $E_{coh}$ highlight the need to develop KEFs which are accurate for both isolated atoms and bulk systems.

## 4. Conclusions

We have presented a functional form and a general fitting procedure for local pseudopotentials, to be used with orbital-free DFT. The pseudopotentials reproduce reference parameters of choice (in our case, lattice parameters and relative energies of several crystalline phases and the bulk modulus) and are matched to a specific OF-DFT setup (i.e. exchange-correlation functional and kinetic energy functional). The compounding of errors of Kohn-Shan DFT and OF-DFT is thereby avoided. We produced local pseudopotentials for Li, Na, and Mg which very accurately reproduce geometries and relative energies of *bcc*, *fcc*, and *hcp* phases. Bulk moduli and vacancy formation energies are reproduced with an accuracy typical of KS-DFT. The fitting procedure is also able to make PPs which reproduce the cohesive energy (which includes single atom calculations) while maintaining acceptable accuracy for bulk properties.

We have provided equations and parameters for the pseudopotentials of Li, Na, and Mg which can be used by others. The approach presented here can be applied to other elements, and availability of local PPs suited for OF-DFT can enable routine large-scale ab initio simulations of many practically relevant materials.

## 5. Acknowledgements

This work was supported by the Tier 1 AcRF Grant by the Ministry of Education of Singapore (R-265-000-494-112).

## 7. Tables

**Table 1.** Lattice parameters (in Å) and differences in energy per atom of different phases of Li (in eV) as well as the bulk modulus $B_0$ (in GPa) and the vacancy formation energy $E_{vac}$ of the *bcc* phase (in eV) obtained with pseudopotentials fitted here (Opt), in comparison with an available PP by Carter and with experimental and highly accurate ab initio (FHI-AIMS) data. Final fitted PP parameters are also given.

| Li | $a_{bcc}$ | $a_{fcc}$ | $a_{hpc}$ | $c_{hpc}$ | $E_{fcc}$-$E_{bcc}$ | $E_{hpc}$-$E_{bcc}$ | $B_0$ | $E_{vac}$ |
|---|---|---|---|---|---|---|---|---|
| Opt WT | 3.44 | 4.33 | 3.06 | 5.00 | -0.0009 | -0.0010 | 15.2 | 0.48 |
| Opt WGC | 3.44 | 4.33 | 3.06 | 5.00 | -0.0009 | -0.0009 | 15.2 | 0.45 |
| Carter[c] WT | 3.39 | 4.26 | 3.01 | 4.93 | -0.0008 | -0.0006 | 16.5 | 0.70 |
| Carter[c] WGC | 3.39 | 4.26 | 3.01 | 4.93 | -0.0008 | -0.0005 | 16.2 | 0.69 |
| FHI-AIMS | 3.44 | 4.32 | 3.06 | 4.99 | -0.0016 | -0.0009 | | |
| Exp. 0 K | 3.45[a] | | | | | | 13.9[a] | |
| Exp. RT | 3.51[b] | | | | | | | 0.48[b] |

| | | | | | Final PP parameters | | | | | |
|---|---|---|---|---|---|---|---|---|---|---|
| $B$ | $b$ | $\beta$ | $C$ | $r_c$ | $c$ | $\gamma$ | $D$ | $r_d$ | $d$ | $\delta$ |
| 5.43191 | 0.60487 | 2.50278 | 0.60374 | 0.90443 | 0.88558 | 2.14722 | -0.14737 | 2.10114 | 0.53573 | 1.50275 |

[a] Ref. [35]
[b] Ref. [36]
[c] http://www.princeton.edu/carter/research/local-pseudopotentials/

**Table 2**. Lattice parameters (in Å) and differences in energy per atom of different phases of Na (in eV) as well as the bulk modulus $B_0$ (in GPa) and the vacancy formation energy $E_{vac}$ of the *bcc* phase (in eV) obtained with pseudopotentials fitted here (Opt), in comparison with experimental and highly accurate ab initio (FHI-AIMS) data. Final fitted PP parameters are also given.

| Na | $a_{bcc}$ | $a_{fcc}$ | $a_{hpc}$ | $c_{hpc}$ | $E_{bcc}$-$E_{hpc}$ | $E_{bcc}$-$E_{fcc}$ | $B_0$ | $E_{vac}$ |
|---|---|---|---|---|---|---|---|---|
| Opt WT | 4.20 | 5.30 | 3.75 | 6.10 | -0.0008 | -0.0009 | 7.4 | 0.32 |
| Opt WGC | 4.20 | 5.30 | 3.75 | 6.10 | -0.0008 | -0.0010 | 7.4 | 0.30 |
| FHI-AIMS | 4.20 | 5.30 | 3.76 | 6.07 | -0.0007 | -0.0013 | | |
| Exp. 0 K | 4.21[a] | | | | | | 7.7[a] | |
| Exp. RT | 4.29[b] | 5.41[b] | | | | | | 0.335[b] |

| Final PP parameters | | | | | | | | | | |
|---|---|---|---|---|---|---|---|---|---|---|
| $B$ | $B$ | $\beta$ | $C$ | $r_c$ | $c$ | $\gamma$ | $D$ | $r_d$ | $d$ | $\delta$ |
| 5.38628 | 0.53765 | 2.12950 | 0.38125 | 1.25984 | 0.80725 | 1.72882 | -0.03790 | 1.94885 | 0.71229 | 1.91424 |

[a] Ref. [35]
[b] Ref. [37]

**Table 3**. Lattice parameters (in Å) and differences in energy per atom of different phases of Mg (in eV) as well as the bulk modulus $B_0$ (in GPa), the vacancy formation energy $E_{vac}$, and the cohesive energy $E_{coh}$ of the *hcp* phase (in eV) obtained with pseudopotentials fitted here (Opt), in comparison with an available PP by Carter [13] and with experimental and highly accurate ab initio (FHI-AIMS) data. Final fitted PP parameters are also given for fits with (Opt$^{coh}$) and without the cohesive energy.

| Mg | $a_{bcc}$ | $a_{fcc}$ | $a_{hpc}$ | $c_{hpc}$ | $E_{fcc}$-$E_{hpc}$ | $E_{bcc}$-$E_{hpc}$ | $B_0$ | $E_{vac}$ | $E_{coh}$ |
|---|---|---|---|---|---|---|---|---|---|
| Opt$^{coh}$ WT | 3.58 | 4.54 | 3.20 | 5.18 | 0.0533 | 0.0483 | 46.2 | 1.10 | 1.53 |
| Opt WT | 3.58 | 4.52 | 3.19 | 5.22 | 0.0087 | 0.0290 | 36.2 | 0.96 | -0.18 |
| Opt WGC | 3.58 | 4.52 | 3.19 | 5.21 | 0.0058 | 0.0229 | 35.9 | 0.78 | |
| Carter WT | 3.58 | 4.53 | 3.19 | 5.22 | 0.0106 | 0.0271 | 37.8 | 1.05 | 0.05 |
| Carter WGC | 3.58 | 4.53 | 3.20 | 5.22 | 0.0068 | 0.0200 | 37.3 | 0.91 | |
| FHI-AIMS | 3.58 | 4.53 | 3.21 | 5.16 | -0.0005 | 0.0007 | | | |
| Exp. 0 K | | | | | | | | | 1.54$^c$ |
| Exp. RT | | | 3.21$^b$ | 5.22$^b$ | | | 35.4$^b$ | 0.58-0.90$^{a,b}$ | |

| Final PP parameters ||||||||||
|---|---|---|---|---|---|---|---|---|---|
| $B$ | $b$ | $\beta$ | $C$ | $r_c$ | $C$ | $\gamma$ | $D$ | $r_d$ | $d$ | $\delta$ |
| 3.33815 | 1.05020 | 2.16052 | -0.10152 | 1.58764 | 1.21079 | 1.85602 | 0.04763 | 2.49299 | 0.65389 | 1.88900 |

| Final PP parameters $E_{coh}$ ||||||||||
|---|---|---|---|---|---|---|---|---|---|
| $B$ | $b$ | $\beta$ | $C$ | $r_c$ | $C$ | $\gamma$ | $D$ | $r_d$ | $d$ | $\delta$ |
| 3.82045 | 1.31633 | 2.17316 | -0.19641 | 2.30156 | 0.61347 | 1.79411 | 0.00082 | 1.83495 | 0.74887 | 1.86846 |

$^a$ Ref. [13]
$^b$ Ref. [38]
$^c$ Ref. [43]

## 8. Figures

Figure 1. The pseudopotentials of Li, Na, and Mg obtained here (Opt). For Li and Mg, the available PPs of Carter [13] are also plotted as dotted curves. For Mg, the PP fitted to $E_{coh}$ is also shown as a grey curve.

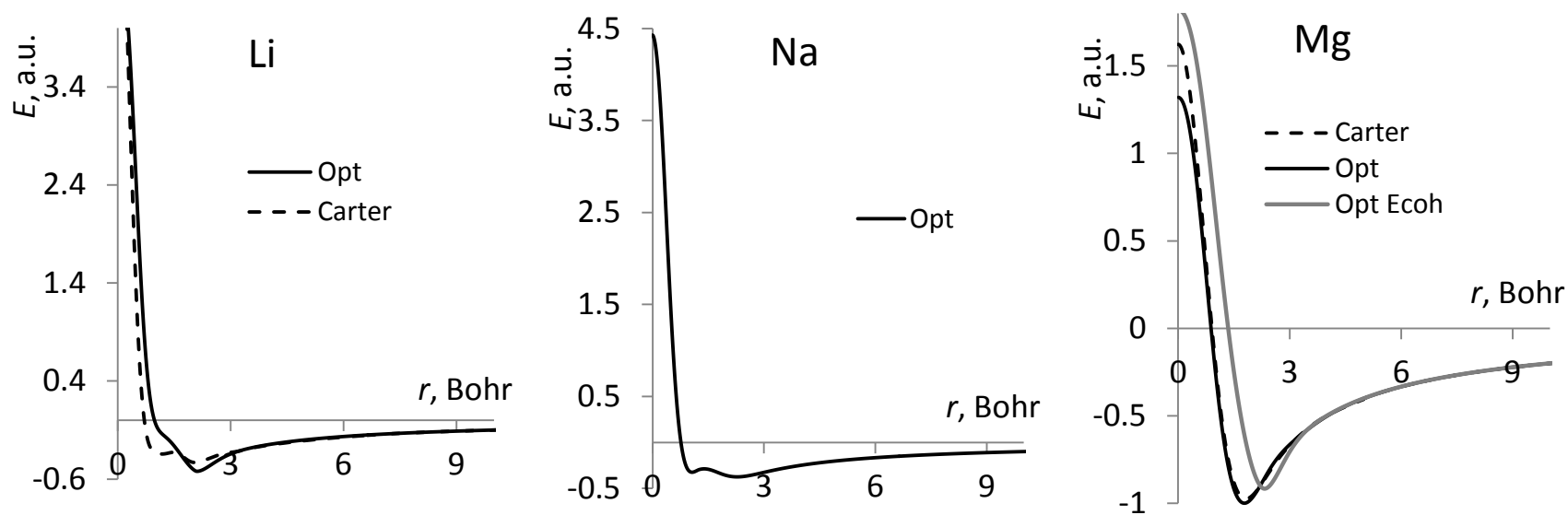